\def\year{2018}\relax
\documentclass[letterpaper]{article} %DO NOT CHANGE THIS
\usepackage{aaai18}  %Required
\usepackage{times}  %Required
\usepackage{helvet}  %Required
\usepackage{courier}  %Required
\usepackage{url}  %Required
\usepackage{graphicx}  %Required
\begin{document}
\title{Collective Story Writing through Linking Images}
\author{Auroshikha Mandal, Mehul Agarwal and Malay Bhattacharyya\\
Department of Information Technology\\
Indian Institute of Engineering Science and Technology, Shibpur\\
Howrah -- 711103, India\\
E-mail: malaybhattacharyya@it.iiests.ac.in
}
\maketitle

\begin{abstract}
Collaborative creativity is the approach of employing crowd to accomplish creative tasks. In this paper, we present a collaborative crowdsourcing platform for writing stories by means of connecting a series of `images'. These connected images are termed as Image Chains, reflecting successive scenarios. Users can either start or extend an Image Chain by uploading their own image or choosing from the available ones. These users are allowed to pen their stories from the Image Chains. Finally, stories get published based on the number of votes obtained. This provides an organized framework of story writing unlike most of the state-of-the-art collaborative editing platforms. Our experiments on 25 contributors highlight their interest in growing shorter Image Chains but voting longer Image Chains.
\end{abstract}

\section{Introduction}
Crowdsourcing involves using the power of crowd to perform a task \cite{brabham2008crowdsourcing}. The sheer power in involving the mass to distribute a job of big proportions makes this idea successful in performing various kinds of tasks, skilled or not-so skilled, technical or creative \cite{kittur2013future}. The aim of this study is to exercise the power of crowdsourcing for carrying out tasks like collaborative story writing, and creative plot building, a field which can not be automated by machines. As the people fill in their text descriptions to make stories, we intend to record the input in the form of creative links between story elements in the form of images (depicting scenarios). Like any crowdsourcing platforms, this too thrives on the abundance of data. As the number of people interacting with the interface increases, the accuracy, diversity and content on the platform also rises. To employ this idea for creative plot building, we have primarily studied the existing collaborative editors and gained insights. This is finally used to design a platform that provides an image based interaction. The stories are basically written through connecting images, termed as Image Chains. This creates a universal platform to merge together ideas of different crowd workers. It has the capability to create growing and evolving stories with time involving increased number of users and is, thus, a step toward organized story writing.

\section{Related Work}
Creative Crowdsourcing is currently a highly exercised concept, with many small start-ups using it to accomplish tasks and attract users. Platforms like DesignHill \cite{designhill} exploit the inputs from crowd workers to help design logos for postings made by people. Another popular platform SquadHelp \cite{squadhelp} employs crowdsourcing to name products and ideas. Graphic designing is also done using crowd inputs by the platforms like 99designs.com \cite{99designs}. However, these platforms work by selecting only one from multiple inputs provided by the crowd contributors. They essentially pick the best out of a pool, with the crowd helping to fill that pool.

CorpWiki is a self-regulating wiki system for effective acquisition of high-quality knowledge content from the corporate employees \cite{Lykourentzou2010}. However, such platforms are not for creative tasks. Collabowriters is a platform that turns crowdsourced inputs into novels. People are allowed to enter lines consisting of a maximum 140 characters, and they are subsequently voted to decide the most popular one. The highest voted line is then added as the next sentence to a novel \cite{collabowriters}. This short lived project has tried to build a well written, coherent story of the size of a book, with the help of crowd. However, we aim at making short stories at first, with the idea being to link creative thoughts together. There are also some Wiki-based interfaces for collaborative story writing. One such platform asks students to edit on a common platform, with an interface like Microsoft Word, and builds new stories through posting a discussion \cite{lspseminar}. The users of this platform have reported that the interface is not receptive to multiple people editing a document simultaneously. This platform also suffers from the problem of content deletion by the other users whenever a new story is being formed. The users have also noticed the lack of an interactive way to add ideas to a story. Some platforms \cite{storybird}, \cite{inklewriter}, \cite{wattpad} allow users to write a complete story online on a platform, such that people can view their stories. An online audience provides continuous feedback to the writers, helping them guide the story, and also to improve the content. This in turn also provides readers with a place to read new stories written by crowd workers. However, these sites work on adding complete stories, and are focused more toward an online platform to judge and read new stories. We merge the working principles of platforms like \cite{storybird} and \cite{lspseminar} to provide users a place to get linked as well as vote for new stories. It does not rely on people being expert story tellers, because people have multiple roles to fulfill. So, most of these said approaches are unorganized.

There are several crowd-powered models that serve the purpose of organized creative writing too. Motif is a recent platform which guides users through adding video snippets from a journey or incident, and adding story-like descriptions to each `scene' they add \cite{kim2015motif}. These are joined together to form coherent stories. Motif thus generates good quality stories by inputs from novices and experts alike, by providing an organized platform for creation. Another platform by Kim et al., Storia \cite{kim2016storia}, works to link social media updates about an event to make a coherent story about a particular incident. The motive remains linking social media updates, but the approach involves asking the crowd to generate summaries from inputs. Storia hence takes the short social media updates from Twitter, Facebook, etc. as nodes in the story, which are to be linked to form a well written story. A crowd-powered model by Kim et al., Mechanical Novel \cite{kim2018mechanical}, attempts at microtasking the 2 facets of story-writing, choosing the target for a story and writing independent scenes of the story, through mTurk. This paper is focused on using the crowd to break down a high-level goal such as creating a story into microtasks which can be self-managed by the crowd to fulfill/extend the primary goal. It allows the crowd to decide on the current state of a story, and how it can be improved or added to. Then the crowd workers propose the changes which should be made to a story, and these changes are voted upon by the others. We however, allow people to merge two story paths together, and to branch one story into a completely new one. The addition of text/task of writing text for an Image Chain, which finally becomes a piece of story, allows people to create what they feel is the best narrative for a given set of images. These story pieces are voted by others to choose the best story for a given sequence. Mechanical Novel does not allow users to continue a current story in a direction they want to, unless the whole crowd decides on it. Our platform aims to provide the flexibility of growing stories in any manner as users want.

\section{Motivational Insights}
The current paper basically aims at building a platform which allows users to add content to a creative story. There are several reasons why a common document editing platform (e.g., Google Doc) will not serve the said purpose. The human co-ordination can be managed by many existing platforms but the challenges remain to be the lack of an organized structure, possible inclusion of noise, chaotic editing, inconsistent results, etc. The main challenges that we have observed are listed below.

\begin{itemize}
\item \textbf{Lack of organization:} If the platform is just an open document which everyone could edit, there is a lot of chaotic input.
\item \textbf{Preservation of content:} People can even delete each others' inputs, and a lot of good ideas get wasted as a result (log files can be ignored by others in the long run).
\item \textbf{Absence of role distribution:} If there is no distribution of roles among the people, it leads to people overriding each others' functions at any instant.
\item \textbf{Recency bias:} Recent edits get more priority than the older edits.
\item \textbf{Arguments related to ownership and content deletion}: Since people can delete others' inputs, it leads to unnecessary arguments between the collaborators.
\end{itemize}

\section{Platform Design}
A text-only platform initially seems like a good idea to connect plots with the help of a crowd. However, as the size of a story increases (the number of scenarios added to a story increases), the complexity of reading through already existing story elements to decide which ones to connect becomes higher. The lack of images makes it difficult for people to easily visualize what other people are creating, without having to read through the whole paragraph. This gives rise to the idea of an even more organized approach, and the ability to interact better with users. We have already pointed out several limitations of existing approaches that led to the better design of our platform.

\begin{figure}
    \centering
    \includegraphics[scale=0.3]{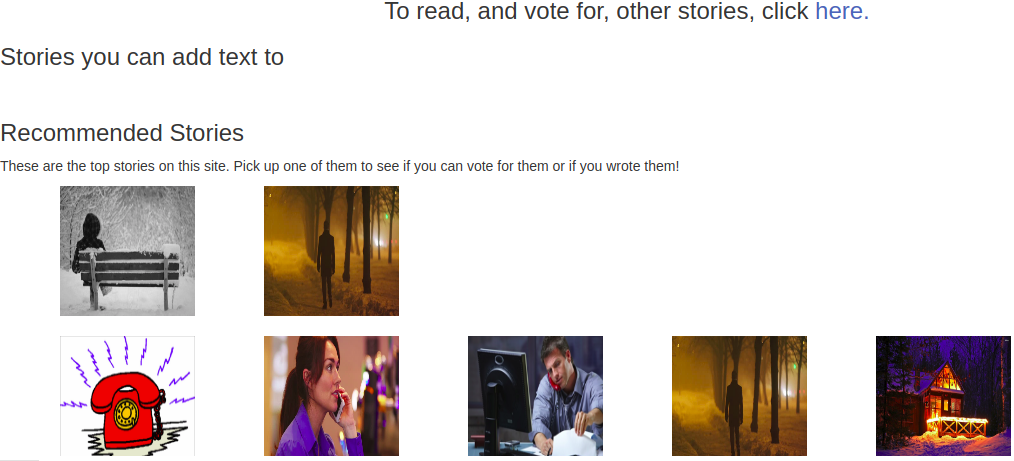}
    \caption{A snapshot of the page where people write their stories or vote for other stories.}
    \label{Fig:Publish}
\end{figure}

In our designed platform, a sequence of images which depicts a flow of narrative is defined as an Image Chain. A starting image from which such Image Chains are formed is referred to as a \textbf{Base Image}. A crowd worker can either start such a story (with a Base Image), continue a story (by extending an Image Chain), or write or edit a story (on an existing Image Chain), and finally vote for such a story (see Fig.~\ref{Fig:Publish}). All these steps, as listed hereunder, are however optional.

\begin{enumerate}
\item \textbf{Starting a story:} A crowd worker can start a story by uploading an image with a description of it or by choosing an image already existing in the database.
\item \textbf{Continuing a story:} A crowd worker can continue a story by selecting a particular Image Chain (ordered chain of images depicting a flow of events created by a crowd worker). Note that, the selected Image Chain also starts with a Base Image. The crowd worker can either upload an (or multiple) image(s) to continue or select an image from the existing database of images. An uploaded image by a user is added on to our pool of images immediately so that the next crowd worker can use the same as and when required.
\item \textbf{Publishing a story:} A crowd worker is entitled to write a story based on the Image Chain he has formed. Every crowd worker who has contributed to the same Image Chains can write their stories by their own or take help from other contributors. Suppose a particular crowd worker has written something about an Image Chain. Subsequent crowd workers writing for the same Image Chain can view what the former has written and use the insights to create another version of the story. Now, the former crowd worker can again use the insights of the latter to create a revised version of the story.
\item \textbf{Voting for a story:} A crowd worker can select a particular Image Chain to vote from the set of all the story chains formed till then. He can select from all the stories written for that Image Chain and vote for his favorite. In this way, a story is voted upon. Internally, votes for an Image Chain are assessed when a crowd worker creates an Image Chain already created by another crowd worker. In that case, instead of creating a redundant Image Chain, we increase the number of votes for that Image Chain. Image Chains with higher votes have a greater probability to be included in the recommendation list.
\end{enumerate}

\section{Empirical Analysis}
Total 25 crowd workers (male = 16, female = 9, mean age = 21.8 years) have taken part in the deployment session by getting connected with computers and mobile phones. None of them are by profession story writers or storytellers. Most of these people have used crowdsourcing platforms earlier, albeit not knowing it is crowdsourced. They have used the platform for 10-72 min (mean time of use = 45 min) in total. During this time, they have used the platform to add images, build stories, and also give feedback about the use, interface and interest via a feedback form. From a starting pool of 30 images (provided as Base Images), the platform has finally grown to 64 images at the end of experimental period of about a month. Total 34 Image Chains (images selected by the crowd workers depicting an ordered flow of thoughts) have been formed and the users have contributed to 22 independent Story Texts.

We have analyzed the Image Chains to study their average length (number of images they contain), and how likely people are to extend chains of a particular length. The average length of an Image Chain is found to be 4.67 after the experimental session, the maximum length being 11. To get an idea about whether users prefer to add images to (extend) smaller chains or bigger ones, we have divided these Image Chains into two groups based on a length threshold value of 5 (Since our average length was calculated as 4.67). Out of these groups, the average number of chains for lengths below ($<=$) 5 images is found to be 5.5 and for lengths greater than 5 images is found to be 2. Putting these two populations under a t-test, we found them to be significantly different from each other ($p$-value = 0.0086; t-test). A possible reason could be that the majority of crowd workers have extended 1-3 sized Image Chains and added 2-3 images more. Hence, even when a crowd worker is adding images to an Image Chain of size $> 7$, the inclination is to extend it to 1-2 images more. Then these crowd workers would end the chain and start writing a story for the same.

\begin{table}
\scriptsize{
    \caption{Analysis of the length of Image Chains and votes obtained by them.}
    \centering
        \begin{tabular}{|c|c|}
        \hline
        Average length of Image Chains & 4.67 \\\hline
        Average number of Image chains of length $\leq 5$ & 5.5 \\\hline
        Average number of Image Chains of length $> 5$ & 2 \\\hline
        Average number of votes for a story text & 3.18 \\\hline
        Average votes for story texts for Image Chains of length $\leq 5$ & 2.4 \\\hline
        Average votes for story texts for Image Chains of length $> 5$ & 3.833 \\\hline
    \end{tabular}
    \label{Table:Length}
}
\end{table}

To ensure whether larger Image Chains obtain more votes, we again compare the two groups of Image Chains (as listed above, with threshold for chain length = 5). The mean and standard deviation values of votes obtained from the users for Image Chains are reported in Table~\ref{Table:Length}. The comparison of the two groups of Image Chains (segregated on the basis of their lengths) was put to a t-test, which gives a significant observation that longer Image Chains obtain significantly higher number of votes ($p$-value = 0.0366; t-test). Hence people are more inclined to alter or grow Image Chains of shorter length ($\leq 5$, in our data), which gives us increased concentration at the lower lengths, while  people are opting to vote more for images of longer lengths, may be because they appear to be more complete as a story.

\section{Conclusion}
Content filtering is one of the primary concerns of a crowdsourcing platform. A system to filter the content, as well as activity, on the platform needs to be present to ensure the quality. The proposed platform attempts to do that by majority approval and storing many versions of one Image Chain with the argument that any chain can be extended later. Additionally, the recommendation facility should be tuned to the genre interest of the user. For this, image descriptions have to be categorized into buckets of similar tastes so that a user selecting images from one bucket is shown images and Image Chains pertaining to the same or similar buckets (buckets with similar kind of genres). Recommendation can also be provided based on the nature of contributions. A crowd worker may extend the work of another crowd worker. Till now, the recommendation facility gives importance only to the voting procedure. Highly voted Image Chains and their corresponding texts are shown in the recommendation section. A balance of votes, genres, contributors and the submission time of any story should make a much better recommendation system. Better incentives are also an important concern here. We did not use any means to incentivize the crowd workers except from providing an encouragement through a Leaderboard. Any such platform would need some form of fund generation or fund collection mechanism to financially support the competent crowd workers.

\section{Acknowledgement}
This publication is an outcome of the R\&D work undertaken in the project under the Visvesvaraya PhD Scheme of Ministry of Electronics \& Information Technology, Government of India, being implemented by Digital India Corporation (formerly Media Lab Asia).

\bibliographystyle{aaai}
\bibliography{Reference}

\end{document}